\documentclass[pra,aps,reprint,citeautoscript,amsmath,amssymb,superscriptaddress]{revtex4-1}
\usepackage{graphicx}
\usepackage{dcolumn}
\usepackage{bm}
\usepackage{amsmath}
\usepackage{hyperref} 
\usepackage{soul} 
\DeclareGraphicsExtensions{.pdf,.eps,.png,.jpg,.mps}
\usepackage{mathrsfs}
\usepackage{soul} 
\usepackage[utf8]{inputenc} 
\usepackage{hyperref} 
\usepackage{xcolor} 
\usepackage{eucal}
\usepackage{setspace}
\usepackage{pifont}

\usepackage{booktabs}

\begin{document}
\title{Entropy and Seebeck signals meet on the edges}

\author{Natalia Cort\'es} 
\email{nc747821@ohio.edu}
\affiliation{Instituto de Alta Investigaci\'on, Universidad de Tarapac\'a, Casilla 7D, Arica, Chile}
\affiliation{Department of Physics and Astronomy, and Nanoscale and Quantum Phenomena Institute, Ohio University, Athens, Ohio 45701–2979, USA}
\author{Patricio Vargas}
\affiliation{Departamento de F\'isica, CEDENNA, Universidad T\'ecnica Federico Santa Mar\'ia, Casilla 110V, Valparaíso, Chile}
\author{S. E. Ulloa}
\affiliation{Department of Physics and Astronomy, and Nanoscale and Quantum Phenomena Institute, Ohio University, Athens, Ohio 45701–2979, USA}

\date{\today}


\begin{abstract}
We explore the electronic entropy per particle $s$ and Seebeck coefficient $\mathcal{S}$ in zigzag graphene ribbons. Pristine and edge-doped ribbons are considered using tight-binding models to inspect the role of edge states in the observed thermal transport properties. As a bandgap opens when the ribbons are doped at one or both edges, due to asymmetric edge potentials, we find that $s$ and $\mathcal{S}$ signals are closely related to each other: both develop sharp dip-peak lineshapes as the chemical potential lies in the gap, while the ratio $s/\mathcal{S}$ exhibits a near constant value equal to the elementary charge $e$ at low temperatures. 
This constant ratio suggests that $\mathcal{S}$ can be seen as the transport differential entropy per charge, as suggested by some authors.  Our calculations also 
indicate that measurement of $s$ and $\mathcal{S}$ may be useful as a spectroscopic probe of different electronic energy scales involved in such quantities in gapped materials.  
\end{abstract}

\maketitle
\date{Today} 
\textit{Introduction}. Nearly forty years ago, Rockwood argued that the thermoelectric power (TEP) or Seebeck coefficient $\mathcal{S}$ in any material is proportional to the total electronic entropy $S$ as $\mathcal{S}=-S/F$   with $F$ the Faraday constant \cite{faraday}. Another relation, the Kelvin formula, connects the TEP with the charge carrier number $N$ derivative of $S$ at constant temperature $T$, $\mathcal{S}_{K}=(1/e)(\partial S/\partial N)_{T}$ \cite{peterson2010kelvin}. It has also been shown that $\mathcal{S} \sim \mathcal{S}_K$ holds qualitatively for noninteracting electrons in a single band (simple metal), in strongly correlated systems \cite{silk2009out}, as well as in the incoherent metal regime in ruthenates  \cite{mravlje2016thermopower}. Several authors have further analyzed the entropy per particle $s=(\partial S/\partial N)_{T}$ to provide a fundamental characterization of the thermodynamics of electronic states in different material systems \cite{entropyperparticledef,varlamov2016quantization,tsaran2017entropy,grassano2018detection,shubnyi2018entropy,galperin2018entropy,sukhenko2020differential,PhysRevB.106.045115}.

The TEP $\mathcal{S}$ is defined as the voltage gradient response $\Delta V$ to a temperature gradient $\Delta T$ at vanishing electric current flux, $\mathcal{S}=\Delta V/\Delta T|_{I=0}$ \cite{mravlje2016thermopower}. $\mathcal{S}$ can be obtained from electronic transport calculations and experiments, revealing characteristic lineshapes as function of gate voltage or chemical potential $\mu$, probing particle-hole asymmetries in the systems \cite{silk2009out}. On graphene-based samples, $\mathcal{S}$ exhibits peak-dip lineshapes given by the contribution of electrons and holes as a gate voltage or $\mu$ changes. For example, in monolayer graphene, a dip-peak curve is seen near the charge neutrality point, broadening its features with increasing temperature \cite{zuev2009thermoelectric}. For gated bilayer graphene systems, the dip-peak shape appears inside the bandgap of the band structure \cite{hao2010thermopower,wang2011enhanced}, and enhanced dip-peak magnitudes are seen when graphene ribbon samples are patterned with defined edges \cite{mazzamuto2011enhanced,hossain2016enhanced}.

\begin{figure}
\includegraphics[width=\linewidth]{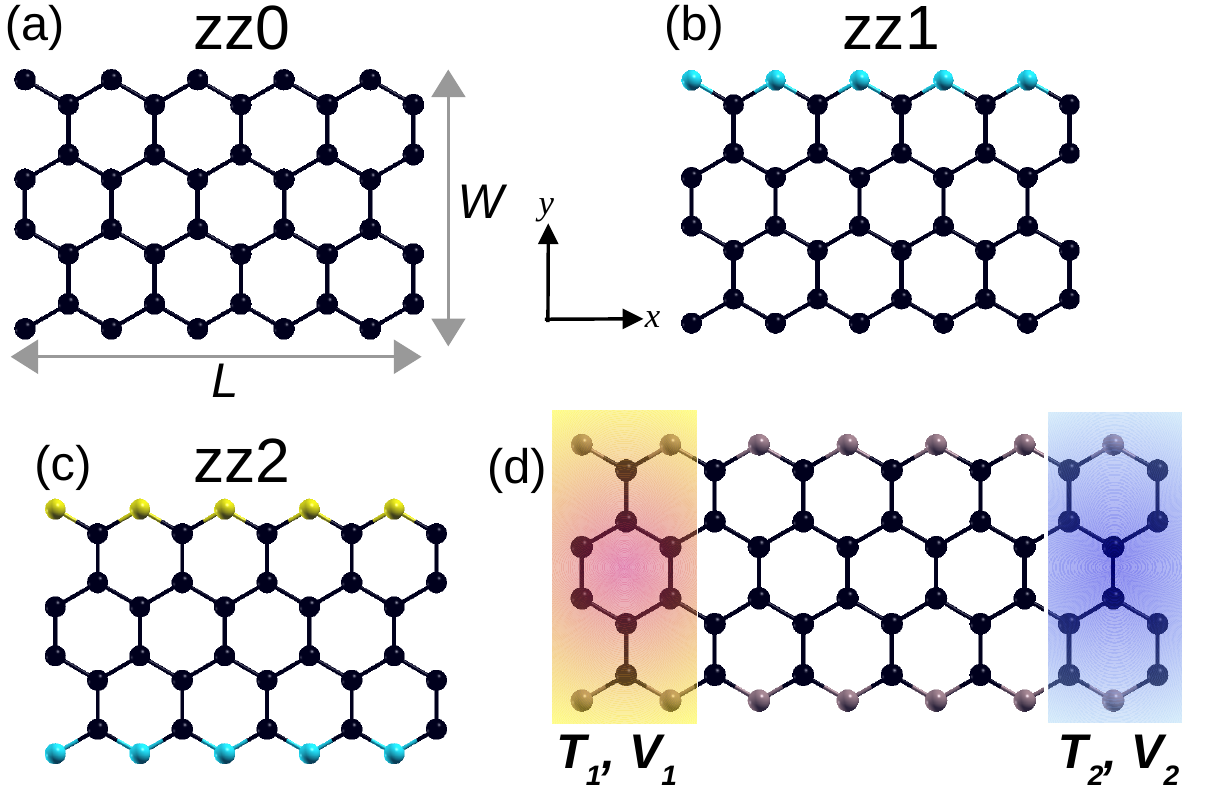}
\caption{Undoped and doped simulated zigzag graphene ribbons. (a) Pristine undoped zz0, (b)  top edge doped zz1, (c)  both edges doped zz2. (d) Each ribbon is connected to two leads, held at voltage $V_1$ and temperature $T_1$ on the left and $V_2$, $T_2$ on the right side. Ribbon size and shape are schematic with length $L$, width $W$, and zigzag edges along the $x$ axis. Black spheres represent carbon atoms with onsite energy $\delta_0=0$, cyan spheres in (b) and (c) stand for edge atoms with onsite energy of $\delta_1=0.2$ eV, yellow spheres in (c) indicate top onsite energies of $\delta_2=-0.4$ eV.}\label{fig1}
\end{figure}

Thermodynamic measurements of the total entropy $S$ (notice $s\neq S$ \cite{entropyperparticledef}) have allowed the acquisition of fundamental information about the electronic state of quantum dots \cite{hartman2018direct}, magic angle twisted bilayer graphene systems \cite{rozen2021entropic,saito2021isospin}, and even a universal value in disordered zigzag graphene ribbons \cite{kim2021topological}. 
Similar analysis of the total entropy $S$ has been carried out in metals and other systems at high temperatures \cite{rinzler2016connecting}, as well as electrons in disordered materials \cite{perez2020entropy}. 
The entropy per particle $s$ provides an excellent thermodynamic tool, exhibiting high sensitivity in low charge density regimes, with experimental evidence of dip-peak curves showing zeros near even filling factors \cite{kuntsevich2015strongly}. Theoretical results for $s(\mu)$ show that it exhibits peak-dip structures in diverse 2D materials, including gapped graphene monolayers \cite{tsaran2017entropy}, semiconducting dichalcogenides \cite{shubnyi2018entropy}, and gated germanene \cite{grassano2018detection}. Although $s$ is not a transport quantity, it can display similar lineshapes as $\mathcal{S}$, suggesting a close interconnection between both quantities not yet explored in 2D materials. We are interested on how these quantities may reflect the spectral features of the system and whether they exhibit similar characteristics in order to fulfill the transport of $s$ as $\mathcal{S}=s/e$ as function of chemical potential and temperature.

To explore these issues, we study the relation between $s$ and $\mathcal{S}$ in zigzag graphene ribbons with flatbands \cite{fujita1996peculiar,nakada1996edge} and gapped states when they are doped on the edges \cite{gunlycke2007altering,moon2016investigations}. The flatbands for zigzag edges near the charge neutrality point (zero energy) are captured by a peak-dip signal in $s$, similar to that produced by the flat state in Lieb's square lattice \cite{PhysRevB.106.045115}. Such flatbands result in vanishing $\mathcal{S}$ values for pristine ribbons but transform into $\mathcal{S}=-2 \frac{k_B}{e} \text{ln}2$ right at the edge state in gapped ribbons. 
Most interestingly, a near proportionality $s \propto \mathcal{S}$ occurs inside the bandgap of doped ribbons, both appearing as sharp peak-dip curves. We in fact find that the ratio $s/\mathcal{S}$ has a nearly constant value of $e$ as $\mu$ shifts within the gap at low $T$, except for a narrow discontinuity around the gap midpoint. We then confirm the near equality $\mathcal{S}\simeq s/e$, demonstrating the Kelvin formula and the relation argued early by Rockwood \cite{rockwood1984relationship}. The relation $s/\mathcal{S}\simeq e$ is expected to be valid for other gapped electronic systems, and their strong sensitivity to gaps and van Hove singularities can be used as practical probes of the electronic structure.   This relation suggests further that the TEP can be seen as the transported entropy per charge, providing an interesting connection between a transport quantity and a thermodynamic measure.

\textit{Model}.
To describe the low-energy spectrum in pristine and edge-doped zigzag graphene ribbons, we use $\pi$-orbital tight-binding models constructed in real space with the pybinding code \cite{dean_moldovan_2020_4010216}. For pristine ribbons [labeled as zz0 in Fig.\ \ref{fig1}(a)], we fix the onsite energies throughout at zero $\delta_0=0$. Edge-doped ribbons are modeled by changing the onsite potentials on the edges.  In a single-edge-doped ribbon [zz1 in Fig.\ \ref{fig1}(b)], the atoms at the top edge have onsite energy $\delta_1=0.2$ eV.\@ For both doped edges (zz2 ribbons), the onsite energies at the top edge are $\delta_2=-0.4$ eV, and $\delta_1$ at the bottom edge, Fig.\ \ref{fig1}(c). 
This allows us to explore the role that edge states have on the thermal transport behavior, which we will see it is quite important, as $s$ and $\mathcal{S}$ show nonequivalent behavior.
With the dispersion relation results for each ribbon system, we calculate the ribbon density of states (DoS), $D$, counting states along the $k_x$-momentum path $X'$-$K$-$\Gamma$-$K$'-$X$. 
The ribbons are then connected to current leads to obtain the charge transport characteristics (transmission probability $\tau$) using the Kwant code \cite{groth2014kwant}. The leads at the ribbon left and right are held at a voltage difference $\Delta V=V_1-V_2>0$, and consider the linear response regime, i.e, $|e \Delta V| \ll \mu $, where $\mu$ is the overall chemical potential. 
To obtain the TEP response, we consider a temperature gradient $\Delta T$ between the leads with $\Delta T=T_1-T_2>0$, as illustrated in Fig.\ \ref{fig1}(d). 
The TEP is quantified by the Seebeck coefficient $\mathcal{S}$, which can be expressed in terms of the thermal integrals $L_n$, as \cite{dollfus2015thermoelectric} 
\begin{eqnarray}
        L_n &=&\frac{2}{h}\int_{-\infty}^{\infty}\tau(\varepsilon)(\varepsilon-\mu)^n\Big(-\frac{\partial f}{\partial \varepsilon}\Big) d\varepsilon,\\
     \mathcal{S} &=&\frac{1}{e T}\frac{L_1}{L_0}=\frac{k_B}{e}\frac{\int_{-\infty}^{\infty}\tau(\varepsilon)\alpha(\varepsilon)\text{cosh}^{-2}\Big(\frac{\alpha(\varepsilon)}{2}\Big)d\varepsilon }{\int_{-\infty}^{\infty}\tau(\varepsilon)\text{cosh}^{-2}\Big(\frac{\alpha(\varepsilon)}{2}\Big)d\varepsilon},\label{seebeck}
\end{eqnarray}
where $h$ is the Plank constant, $k_B$ the Boltzmann constant, $\varepsilon$ the energy eigenvalues for each system, $f(\varepsilon,T,\mu)=1/[e^{\beta(\varepsilon-\mu)}+1]$ the Fermi-Dirac distribution with $\beta=1/k_{B}T$, $\tau(\varepsilon)$ the transmission probability function, and $\alpha(\varepsilon)=(\varepsilon- \mu)/{k_B T}$. Similarly, the entropy per particle $s$ can be expressed as \cite{grassano2018detection,galperin2018entropy,sukhenko2020differential} 
\begin{equation}\label{entropy}
    s=k_B\frac{\int_{-\infty}^{\infty}D(\varepsilon)\alpha(\varepsilon)\text{cosh}^{-2}\Big(\frac{\alpha(\varepsilon)}{2}\Big)d\varepsilon }{\int_{-\infty}^{\infty}D(\varepsilon)\text{cosh}^{-2}\Big(\frac{\alpha(\varepsilon)}{2}\Big)d\varepsilon}.
\end{equation}
Note the similarity of Eqs.\ \ref{seebeck} and \ref{entropy}, considering that by obtaining $\tau(\varepsilon)$ and $D(\varepsilon)$ respectively, one can capture the equivalence and/or difference between $\mathcal{S}$ and $s$, providing an efficient and reliable approach to study thermally-activated electronic signals in diverse quantum materials, such as gapped graphene ribbons.

\begin{figure}[!h]
\centering
\includegraphics[width=\linewidth]{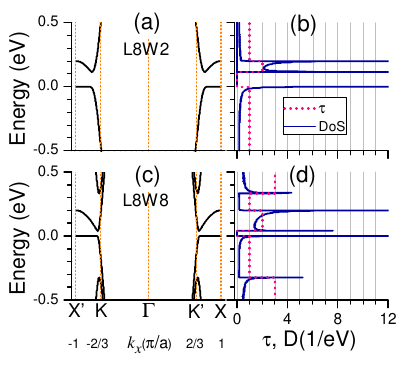}
\caption{Electronic spectra for ribbons zz1. Top row $L8W2$, bottom row $L8W8$. (a), (c) band structure along $X'$-$K$-$\Gamma$-$K$'-$X$; (b), (d) transmission $\tau$ and density of states $D$. Numbers along the horizontal axis indicate high symmetry $k_x$ values, highlighted with orange vertical lines in (a) and (c). Gray vertical lines in (b) and (d) indicate integer steps for $\tau$.}\label{fig2} 
\end{figure}

\textit{Results and discussion.}
We compare graphene ribbons zz1 and zz2 with different sizes and contrast their response with that of pristine ribbons zz0. 
We look at ribbons with a nominal length $L=8$ nm; $L$ is large enough to consider these results valid for any mesoscopic ribbon.  Most relevant characteristic is the ribbon width, as it determines the spacings of the bulk subbands and associated DoS features.
The two ribbon widths considered, $W=2$ nm and $W=8$ nm, which we label $L8W2$ and $L8W8$, respectively. 
It is well known that pristine zigzag ribbons are metallic, regardless of the width, with flat bands at zero energy \cite{supplemental}, due to extended states along both ribbon edges.  Edge-atom doping changes that, opening a gap at the charge neutrality point. Figure \ref{fig2}(a)-(b) show the band structure, $D$ and $\tau$ for a zz1-$L8W2$ ribbon, while (c) and (d) are for the zz1-$L8W8$ ribbon. Both ribbons exhibit gaps with a magnitude of $\delta_1=0.2$ eV at the $X$,$X'$ points, and smaller gaps near the $K$,$K'$ points of $\simeq 0.11$ eV for $L8W2$, and $\simeq 0.04$ eV for $L8W8$. The energy gaps near $K$,$K'$ decrease as the ribbon width increases and the two edges further decouple \cite{son2006energy}. The gaps at $X$,$X'$ open because of the asymmetric onsite potentials on the bottom ($\delta_0=0$) and top ($\delta_1$) ribbon edges, which break inversion symmetry in the ribbon.
%
The top valence band shows a zero energy flatband within the  ($X',K$) and ($K',X$) windows, associated with the unperturbed bottom edge \cite{supplemental}.  The corresponding DoS shows a large peak around zero energy, for both ribbon widths. 
The gap opening near $K,K'$ results in a parabolic bottom conduction band with local inverted curvature at the $X,X'$ points. 
These characteristics result in large van Hove singularities in the DoS at the energies of the bottom conduction and inverted bands.
Additional van Hove peaks at higher energies are due to the onset of bulk subbands, as those shown for $L8W8$ in panel (c) and (d), at energies $\simeq \pm 0.32$ eV. These bulk states at larger energies are common for pristine and doped ribbons \cite{supplemental}. The DoS naturally vanishes for energies within the bandgap near the $K,K'$ points. The gaps and van Hove peaks in the DoS will be shown to produce strong signatures in the entropy per particle response, as anticipated from Eq.\ \ref{entropy}. 

When $\Delta V$ is turned on at $\Delta T=0$, charge carriers can be transported along with allowed zigzag channels in the ribbon system with probability $\tau(\mu)$. In Fig.\ \ref{fig2}(b) and (d), $\tau$ (dotted pink lines) jumps from $1$ to $0$ at the top valence energy, vanishes in the energy gap, and jumps from $0$ to $2$ to $1$ near the bottom of the conduction band. This quantized electronic transport in ribbons is directly linked to $\mathcal{S}$ via Eq.\ \ref{seebeck}. When the ribbons are placed in a temperature gradient $\Delta T$, the charge carriers are thermally excited and move from the hottest to the coldest lead and vice versa.

\begin{figure}[!h]
\centering
\includegraphics[width=\linewidth]{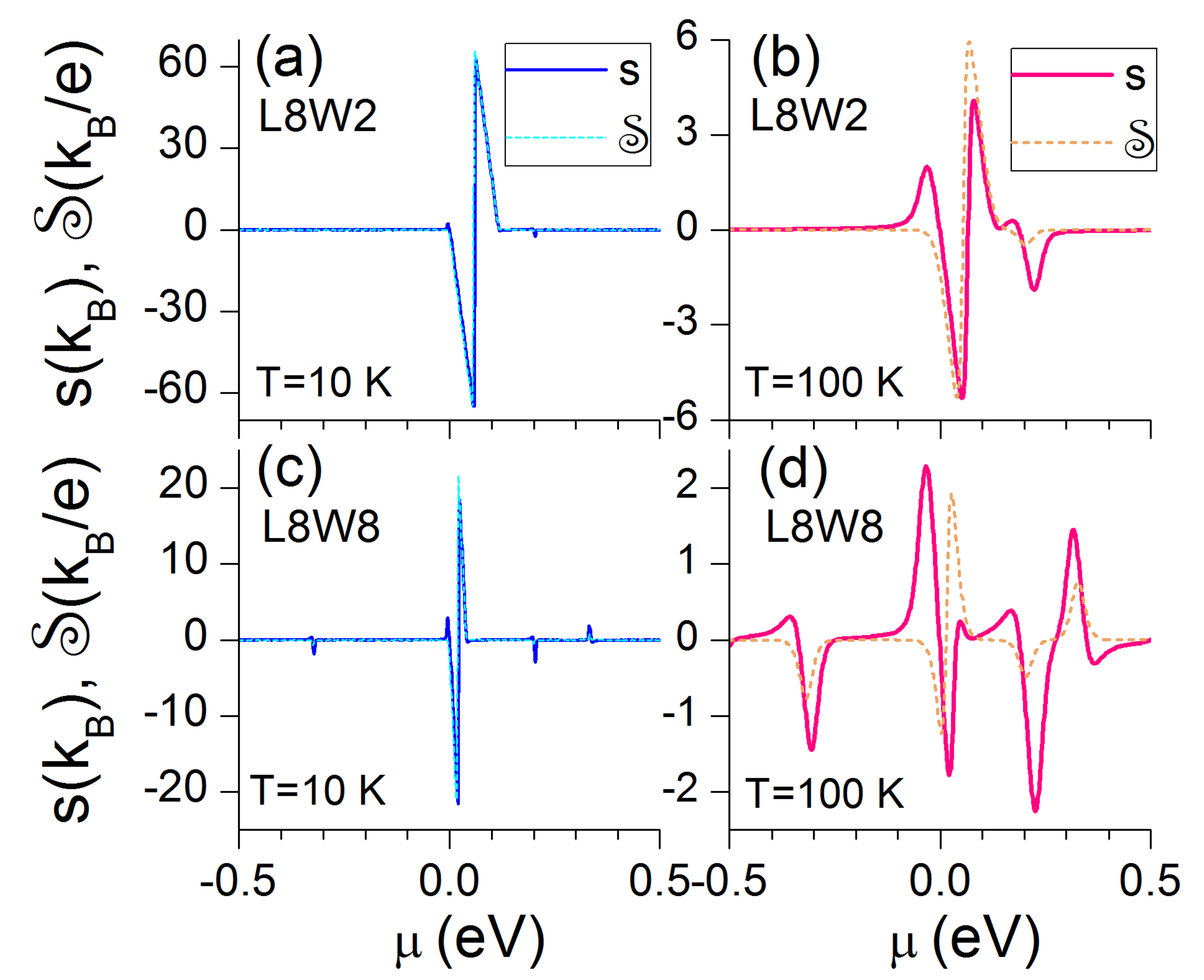}
\caption{Entropy per particle $s$ (solid lines) and Seebeck signal $\mathcal{S}$ (dashed lines) for zz1 ribbons. Top row, narrow ribbon, $L8W2$; bottom row, wide ribbon, $L8W8$. (a), (c) $T=10$ K; (b), (d) $T=100$ K. Notice different vertical scales.}\label{fig3}
\end{figure}

In Fig.\ \ref{fig3} we present $\mathcal{S}$ (dashed) and $s$ (solid lines) signals as function of $\mu$ at $T=10$ and $100$ K for different width zz1 ribbons. The $\mu$ scan can be implemented by gate voltages, which would produce corresponding charge density changes in the system \cite{wang2011enhanced}. At low temperatures, $T=10$ K, Fig.\ \ref{fig3}(a), (c), $s$ and $\mathcal{S}$ exhibit a dip-peak structure with nearly identical shape and amplitude within the energy gap of each ribbon. The shapes are sharper for $L8W8$ as the electronic structure [Fig.\ \ref{fig2}(c)] shows a narrower gap for wider ribbons. The large discontinuous sign change for both $s$ and $\mathcal{S}$ occurs  near the gap midpoint $\mu_{\text{gmp}}$, as the contributions from charge carrier densities (electrons and holes) cancel each other at this $\mu_{\text{gmp}}$ value. The similarity $s\simeq \mathcal{S}$ whenever $\mu$ crosses the gap comes from the vanishing of the relevant quantity,  $\tau(\varepsilon) = D(\varepsilon) = 0$ in this region, making Eqs.\ \ref{seebeck} and \ref{entropy} equivalent. As $T$ increases to 100 K, Fig.\ \ref{fig3}(b) and (d), $s$ and $\mathcal{S}$ decrease by one order of magnitude, broaden their shape and are no longer similar.
Interestingly, the presence of the flat edge state and its associated sharp DoS are captured by a positive peak in $s$ at $\mu \simeq 0$, decreasingly only slightly with $T$.  In contrast, $\mathcal{S}(\mu=0)$ has a finite value $=-2 \frac{k_B}{e}\text{ln}\, 2$ at low temperatures ($\lesssim 100$ K) for both ribbon widths, as expected from an analytical estimate that sets $\tau$ as a Heaviside function \cite{supplemental}.
The (inverted) parabolic band edge at $\varepsilon=0.2$ eV is seen in both $s$ and $\mathcal{S}$ as a negative peak near $\mu=0.2$ eV, with larger amplitude for $s$, that decreases with $T$. 
The flat and parabolic $s$ and $\mathcal{S}$ edge responses will be discussed in more detail below. 

As $\mu$ shifts away from the gap edges for $L8W8$ in Figs.\ \ref{fig3}(c) and \ref{fig3}(d), bulk subband features appear in $s$ and $\mathcal{S}$, with $s$ showing a sign change near each DoS maximum.  The peaks in $\mathcal{S}$ are positive for electrons and negative for holes--such sign reversal is clear in Fig.\ \ref{fig3}(d) for $\mu \sim \pm 0.3$ eV.\@ Similar behavior for $s$ and $\mathcal{S}$ at larger $\mu$ values is also present for bulk subbands in pristine ribbons \cite{supplemental}.  One could use such sign reversal in $\mathcal{S}(\mu)$ to monitor subband curvature changes, as external fields (e.g., strains or voltages) may produce band inversions \cite{AbdulMassInv}.

\begin{figure}[!h]
\centering
\includegraphics[width=\linewidth]{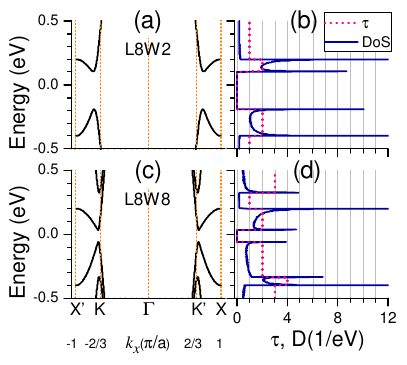}
\caption{Electronic spectra for zz2 ribbons. Top row $L8W2$, bottom $L8W8$. (a), (c) $X'$-$K$-$\Gamma$-$K'$-$X$ band structure; (b), (d) transmission $\tau$ and density of states $D$.}\label{fig4}
\end{figure}

\begin{figure}[!h]
\centering
\includegraphics[width=\linewidth]{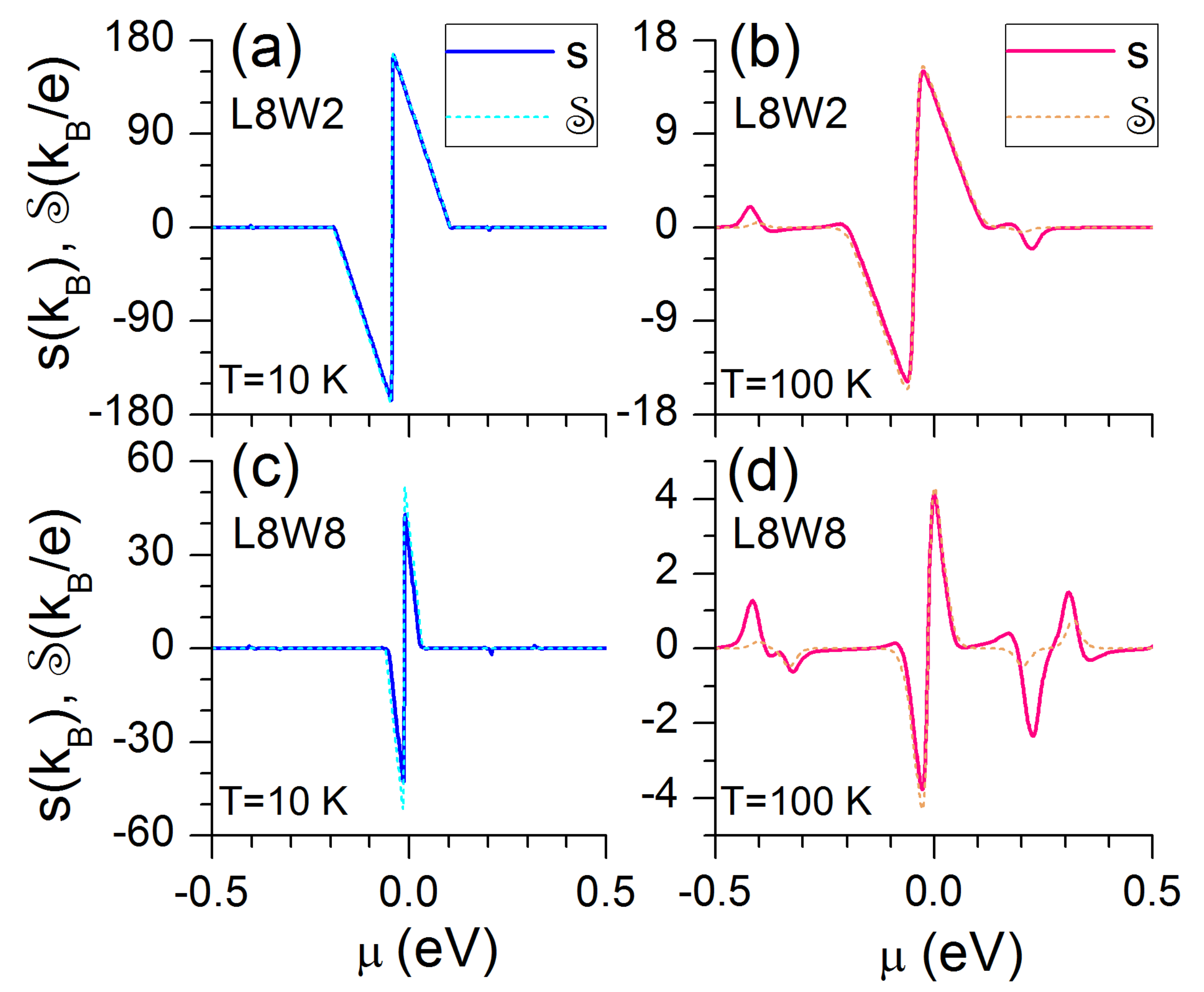}
\caption{Entropy per particle $s$ (solid lines) and Seebeck signal $\mathcal{S}$ (dashed lines) for zz2 ribbons. Top row $L8W2$, bottom row $L8W8$. (a), (c) $T=10$ K; (b), (d) $T=100$ K. Notice different vertical scales.}\label{fig5}
\end{figure}

Another interesting case is when both zigzag edges are asymmetrically doped, as in the zz2 ribbons in Fig.\ \ref{fig1}(c). This system is similar to ribbons with hydrogen-oxygen doped edges \cite{moon2016investigations}, or to a non-magnetic version of ribbons with antiferromagnetic edges \cite{son2006energy}. Figure \ref{fig4} shows the electronic dispersion, DoS and $\tau$ for $L8W2$ and $L8W8$ zz2 systems. The gaps are larger than for zz1 ribbons, as the onsite potentials on the edges contribute additively, producing $0.6$ eV gaps at $X,X'$. Near $K,K'$ the gaps narrow to $\simeq 0.3$ eV for $L8W2$ and $\simeq 0.1$ eV for $L8W8$. We notice there is no flat edge state around zero energy as in zz1 or pristine ribbons.  Instead, there is an asymmetric gap about $\varepsilon=0$, and the edge dispersions are parabolic.  The structure is otherwise similar 
to the case shown in Fig.\ \ref{fig2}, with rescaled energies: $\tau$ and $D$ present similar structure to the zz1 devices but with larger band gaps. 
As a consequence, the dip-peak structure for $s$ and $\mathcal{S}$ in Fig.\ \ref{fig5} is nearly identical within the gaps, with $s\simeq  e \mathcal{S}$ at both $10$ K and $100$ K.\@  The vanishing DoS on both ribbon gap edges results in even more symmetric responses in $s$ and $\mathcal{S}$ for zz2 systems.      

\begin{figure}[!h]
\centering
\includegraphics[width=\linewidth]{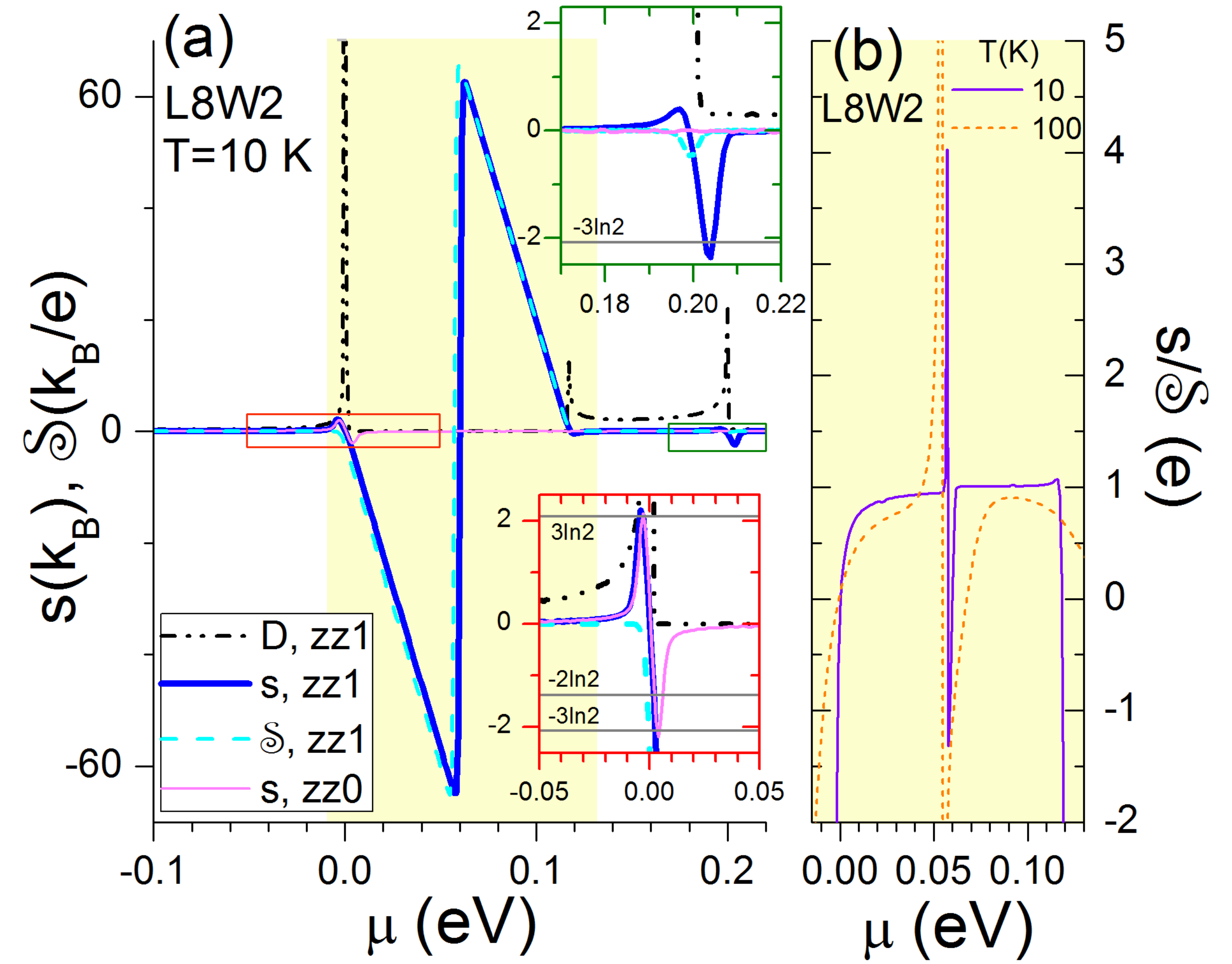}
\caption{(a) Entropy per particle $s(\mu)$ (solid blue) and Seebeck signal $\mathcal{S}(\mu)$ (dashed cyan line) at $T=10$ K for zz1-$L8W2$ ribbon. Dash-dot black line is the density of states $D$; solid magenta line shows $s$ for a pristine zz0-$L8W2$ ribbon. Red (green) rectangle indicates areas near the edge states at $\mu=0$ ($\mu=0.2$ eV) in the bottom (top) inset. (b) Ratio $s(\mu)/\mathcal{S}(\mu)$ in units of $e$ for curves within the yellow rectangle in (a), results for $T=100$ K are also included.}\label{fig6}
\end{figure}

We now turn our attention to the zz1-$L8W2$ ribbon electronic-thermodynamic state [Fig.\ \ref{fig2}(a)-(b) and Fig.\ \ref{fig3}(a)-(b)] as it presents a flat band at the valence gap edge and a parabolic band at the conduction gap edge. Figure \ref{fig6}(a) clearly exhibits the dip-peak shapes of $s$ and $\mathcal{S}$ within the gap at $T=10$ K (region highlighted in yellow); the small mismatch near midgap is related to the asymmetric shape of the DoS (dot-dashed black line), especially around the flatband near $\mu=0$. The area inside the red rectangle focuses on $s$ and $\mathcal{S}$ near the flatband, as shown amplified in the bottom inset. $s$ (blue solid line) presents a positive peak of height $s \sim 3k_B \text{ln}2$ for negative $\mu$, changes sign near $\mu=0$ where $D$ has a maximum, and then continues with a constant slope for positive $\mu$ \cite{supplemental}. The inset also shows $s$ for the flatband of the pristine ribbon (thin magenta line) with an antisymmetric shape around $\mu=0$ with peak values of $s \sim \pm 3k_B \text{ln}2$ \cite{supplemental,entropyvaluesdef}.  In contrast, $\mathcal{S}$ (dashed cyan line) drops monotonically at the gap edge, reaches a  value $\mathcal{S}(\mu=0)=-2\frac{k_B}{e}\text{ln}2$, before having a constant slope for $\mu \lesssim \mu_{\text{gmp}}$  \cite{supplemental}. 

The parabolic band edges at $X$,$X'$ produce features at $\mu \simeq 0.2$ eV, as displayed in the top inset. The associated van Hove peak in $D$ produces a sign change in $s$ with peak value $\sim  -3k_B \text{ln}2$, whereas $\mathcal{S}$ has a negative peak of height $\mathcal{S}\simeq -0.46$ $k_{B}/e$, its sign indicating the inverted parabolic dispersion. Approximate results for $s$, $\mathcal{S}$ arising from edge and bulk states can also be described by Sommerfeld expansions for different $L_n$ integrals \cite{supplemental}, which agree with these results.  

Figure \ref{fig6}(b) shows the ratio $s(\mu)/\mathcal{S}(\mu)$ for the gap region highlighted in yellow in Fig.\ \ref{fig6}(a). The ratio presents an asymmetric lineshape due to the asymmetry of both $\tau$ and $D$ at the gap edges of the zz1-$L8W2$ ribbon, see Fig.\ \ref{fig2}(b). The ratio at $T=10$ K grows from the valence gap edge near $\mu=0$ reaching a constant value of $\sim e$ across the gap, except for a sharp discontinuity at midgap  ($\mu_{\text{gmp}} \simeq 0.055$ eV), and falls down near the conduction gap edge. At higher temperature, $T=100$ K, the ratio $s/\mathcal{S}$ shows smoother variation and nearly constant $e$ value over a smaller region. 
The relation $\mathcal{S} \simeq s/e$ is also valid in the gap region of zz2 ribbons and seen to persist even at higher temperatures, as expected from the larger energy scales involved \cite{supplemental}.
The equivalence between electronic transport and thermodynamic response, as given by $\mathcal{S}\simeq s/e$ suggests that $\mathcal{S}$ can be regarded as the transported entropy per unit charge in the gapped regime. The connection between these quantities could be explored and exploited in different materials. 

\textit{Conclusions}. 
When graphene ribbons are asymmetrically doped along one of the zigzag edges a gap opens, while a flatband remains along the pristine edge of the ribbon. The entropy per particle $s$ is sensitive to the flatband, resulting in an asymmetric peak-dip curve whereas the Seebeck signal $\mathcal{S}$ has a finite value of $\mathcal{S}=-2 \frac{k_B}{e}\text{ln}2$ right at the flatband energy. $s$ and  $\mathcal{S}$ reach their highest amplitudes inside the gap with a dip-peak structure and fulfilling the relation $\mathcal{S}\simeq s/e$ all across the gap--except at midpoint.  This relation is especially clear at low temperatures, since at higher temperatures its dependence on chemical potential is blurred and softened.
The Seebeck coefficient can then be seen as the transported entropy per charge within the gapped regime. 
The large magnitudes of $s$ and $\mathcal{S}$ signals within transport gaps can be useful for bandgap estimation \cite{goldsmid1999estimation}, while the sign $\mathcal{S}$ is determined by the local band curvature. 
It would also be interesting to explore if the ratio 
$s/\mathcal{S}$ as function of chemical potential can indicate changes in the quasiparticle charge as materials may undergo transitions to strongly correlated regimes and possible charge fractionalization \cite{Senthil2002}.

\textit{Acknowledgments.}
N.C. acknowledges support from ANID Fondecyt Iniciaci\'on en Investigaci\'on No. 11221088 and IAI-UTA, and the hospitality of Ohio University, P.V. acknowledges support from ANID Fondecyt Regular No. 1210312, and S.E.U. acknowledges support from U.S. Department of Energy, Office of Basic Energy Sciences, Materials Science and Engineering Division.

\bibliography{bib}

\begin{thebibliography}{38}
\bibitem [{far()}]{faraday}%
  \BibitemOpen
  \href@noop {} {}\bibinfo {howpublished} {The Faraday constant is the product
  of the elementary charge $e$ and Avogadro number $N_A$, $F \equiv e
  N_A$.}\BibitemShut {Stop}%
\bibitem [{ent({\natexlab{a}})}]{entropyperparticledef}%
  \BibitemOpen
  \href@noop {} {}\bibinfo {howpublished} {The electronic entropy per particle
  $s$ is also referred as differential entropy per particle. Through Maxwell
  relations, $s$ can also be expressed as $s=\Big(\frac{\partial S}{\partial
  N}\Big)_{T}=-\Big(\frac{\partial \mu}{\partial
  T}\Big)_{N}=\Big(\frac{\partial N}{\partial T}\Big)_{\mu}\Big(\frac{\partial
  N}{\partial \mu}\Big)_{T}^{-1}$, with $S$ the total electronic entropy and
  $N=\int_{-\infty}^{\infty}D(\varepsilon)f(\varepsilon,T,\mu)d\varepsilon$ the
  charge carrier number at thermal equilibrium.} ({\natexlab{a}})\BibitemShut
  {NoStop}%
\end{thebibliography}

\end{document}